\documentstyle[aps,multicol,epsf]{revtex}

\draft
\def\squote{}
\def\quote#1#2#3#4{\squote {#1,\ {\sl#2}\ {\bf#3}, #4}.\par}
\def\qquote#1#2#3#4{\squote {#1,\ {\sl#2}\ {\bf#3}, #4};}

\def\prl{{\sl Phys. Rev. Lett.}\ }

\def\lphi{$L_\phi$}

\def\pr {{\sl Phys. Rev.}\ }

\def\e{\epsilon}
\def\deriv{\partial}

\begin{document}
%\tighten
\title{Percolation-type description of the metal-insulator transition in
 two dimensions}
%}}
\bigskip
\author{\large  Yigal Meir}
\address{Department of Physics, Ben-Gurion University, Beer Sheva 84105, ISRAEL}
\maketitle
\begin{abstract}
A simple non-interacting-electron
model,  combining local quantum tunneling and global classical 
percolation
(due to a finite dephasing time at low temperatures),  is introduced to 
describe a metal-insulator transition in two dimensions. It is shown that many
features of the experiments,  such as the exponential dependence of the resistance
on temperature on the metallic side, the linear dependence of the exponent
on density, the $e^2/h$ scale of the critical resistance, 
the quenching of the metallic phase by a parallel magnetic field
and the non-monotonic dependence of the critical density on a perpendicular
 magnetic field,  can be naturally explained by the model.
\end{abstract}
%\newpage
\begin{multicols}{2}
The experimental observation of a metal-insulator transition in
two dimensions
\cite{kravchenko,coleridge,simmons,hanein,yaish}
 has been a subject of extensive investigation, since
it is in disagreement with the predictions of single-parameter
scaling theory for noninteracting electrons \cite{abrahams}.
Several theories, based on the treatment of disorder and
electron-electron interactions  by Finkelstein \cite{finkelstein}, have
been put forward \cite{interactions}.
Other approaches considered spin-orbit scattering \cite{pudalov_model}, 
percolation of electron-hole liquid \cite{he} or scattering by
impurities \cite{altshuler}. 
%(in the last two models the transition is actually
%a finite temperature crossover phenomenon). 
To date there is no acceptable
microscopic theory that describe quantitatively the observed data.

Here we present a simple non-interacting electron model,  combining local
quantum tunneling and global classical percolation,  to explain several 
features of the experimental  observations.
% As the data of the GaAs samples and the Silicon
%samples is different in several aspects,  we concentrate on the former.  
The main observations we want to understand are the following:
\begin{itemize}
\item As the system is cooled down
% there is a critical density,  such that
the resistance of samples with density  higher than some critical density
extrapolates to a finite value
at zero temperature,  while that of samples with lower density diverges.
\item The resistance of the sample with the critical density does not
depend on temperature (at least for a limited range of low temperatures).
\item The conductance of the critical-density sample is of the order of
$e^2/h$.
\item On the metallic side the functional dependence of the resistance is of
 the form $R(T)=R_0+R_1 \exp(-A/T)$. The parameter $A$ 
varies linearly with the density and vanishes at the transition.
\item In perpendicular magnetic fields this transition is continuously connected
 with the quantum Hall--Insulator transition \cite{magfield}. The critical
 density varies nonmonotonically with magnetic field,  with a minimum around 
 $\nu=1$.
\item 
Parallel magnetic fields destroy the metallic phase,  at least 
for densities near the transition \cite{parallel,simmons}.
\end{itemize} 
Before introducing the model let us mention three
 other experimental observations
-  (a) The strong disorder is crucial to see the transition. In GaAs the
transition is seen only at samples with low mobility (even with the same
 density). In fact, Ribeiro et al. \cite{ennslin} have recently observed
  a zero-field metal-insulator transition in high-density
  n-type GaAs sample with strong enough disorder (which was 
introduced by a matrix of randomly distributed quantum dots). (b) There is
additional experimental indications that the transition is not driven
by interactions -- Yaish and Sivan \cite{yaish} studied a system of two
 parallel gases,  one of electrons and one of holes. The observed 
 metal-insulator transition
 in the hole gas depended only slightly on the electron density,  even though
 one expects that increasing electron density will screen the interactions
 between holes and suppress the metallic phase in the hole gas. On the 
 contrary, 
 increasing  electron density led to increasing conductance in the hole gas, 
 indicating that its main role is to screen the impurity potentials in the
 hole gas.
(c) There is
a growing experimental evidence that even at the lowest available temperatures, 
the dephasing length,  \lphi,  is finite \cite{webb,marcus,joe}.
% This suggests that on
% a length scale larger than \lphi,  transport is classical.

 Based on all these observations we now  suggest the following
 scenario to explain the experimental observations --
  the potential fluctuations due to the 
 disorder define density puddles
 of size \lphi\ or larger in which
 the electron wavefunction totally dephases. (Density separation into puddles
 in gated GaAs was indeed observed experimentally by Eytan et al. 
 \cite{israel},  using near-field spectroscopy.)
 Locally, between these puddles,  
 transport is via  quantum tunneling through saddle points,
  or quantum point-contacts (QPCs). Since between
 such tunneling events dephasing takes place,  the conductance
 of the system will be determined by adding classically these quantum
 resistors. A related model was introduced by Shimshoni et al. \cite{efrat}
 to describe successfully transport in the quantum Hall (QH) regime.
 In fact,  it was later concluded \cite{dephasing} that the observation of
  a finite,  quantized Hall resistance in the Hall insulator phase 
 \cite{qHduality,qHinsulator} can only occur when the dephasing length is 
 smaller than
 the size of these puddles -- otherwise the Hall resistance diverges
 \cite{ora}. In addition,  this model also accounted for the observed 
 current-voltage duality around the transition \cite{qHduality},  
 a duality which
 was also observed in the zero-field transition \cite{duality,coleridge}.
  The percolative
 nature of the system in the QH regime was indeed verified 
 experimentally \cite{qh}.
Finite dephasing length can also explain the observed non-critical behavior of
 the resistance near the QH-insulator transition \cite{noncritical}.
 We will return to the QH regime below.
 
 We characterize each saddle point by its critical energy $\e_c$, such that the
 transmission through it is given by $T(\e)=\Theta(\e-\e_c)$. 
 %(we assume that the
 %energy scale over which the transmission changes from zero to unity is smaller
 %than the other relevant energy scales,  to avoid additional parameters). 
 Thus
 the conductance through each QPC is given by the Landauer formula, 
\begin{eqnarray}
G(\mu,T) &=& {2e^2\over h} \int d\e
 \left(-{{\deriv f_{FD}(\e)}\over{\deriv \e}}\right)
 T(\e)\nonumber\\ 
  &=&  {2e^2\over h} {1\over{1+\exp[(\e_c-\mu)/kT]}} , 
\label{QPC}
\end{eqnarray}
where $\mu$ is the chemical potential,  and
$f_{FD}$ is the Fermi-Dirac distribution function. 

The system is now composed of classical resistors,  where the resistance of each one
 of them is given by (\ref{QPC}),  with random QPC energies.
 In the numerical data presented below,  
 we solved  a 20x20 system of QPCs (which,  for simplicity, 
 has the topology of a square lattice),  each averaged over 1000 
 realizations of disorder, where  the 
 QPC energies were taken from a square distribution of width $W$.
At zero temperature the conductors have either zero conductance or a conductance
 equal to $2e^2/h$ and one has the usual second-order percolation transition.
 The critical conductance exponent $t$ is known at two dimensions and is equal
 to $1.3$ \cite{stauffer}. In Fig.~1 we fit the experimental data of
 \cite{hanein}  and of \cite{yaish} to the
 expected critical  dependence.
 Clearly, the agreement with the classical percolation  prediction is excellent.
Moreover, the fact that while the density scale is so different between the
two experiments, the conductance scale is identical, clearly demonstrates
that $e^2/h$ is the only conductance scale in the system.
\vskip  -0.5 truecm
\begin{center}
\leavevmode \epsfxsize=3.5in
%\epsfbox{dft_fig.ps}
\epsfbox{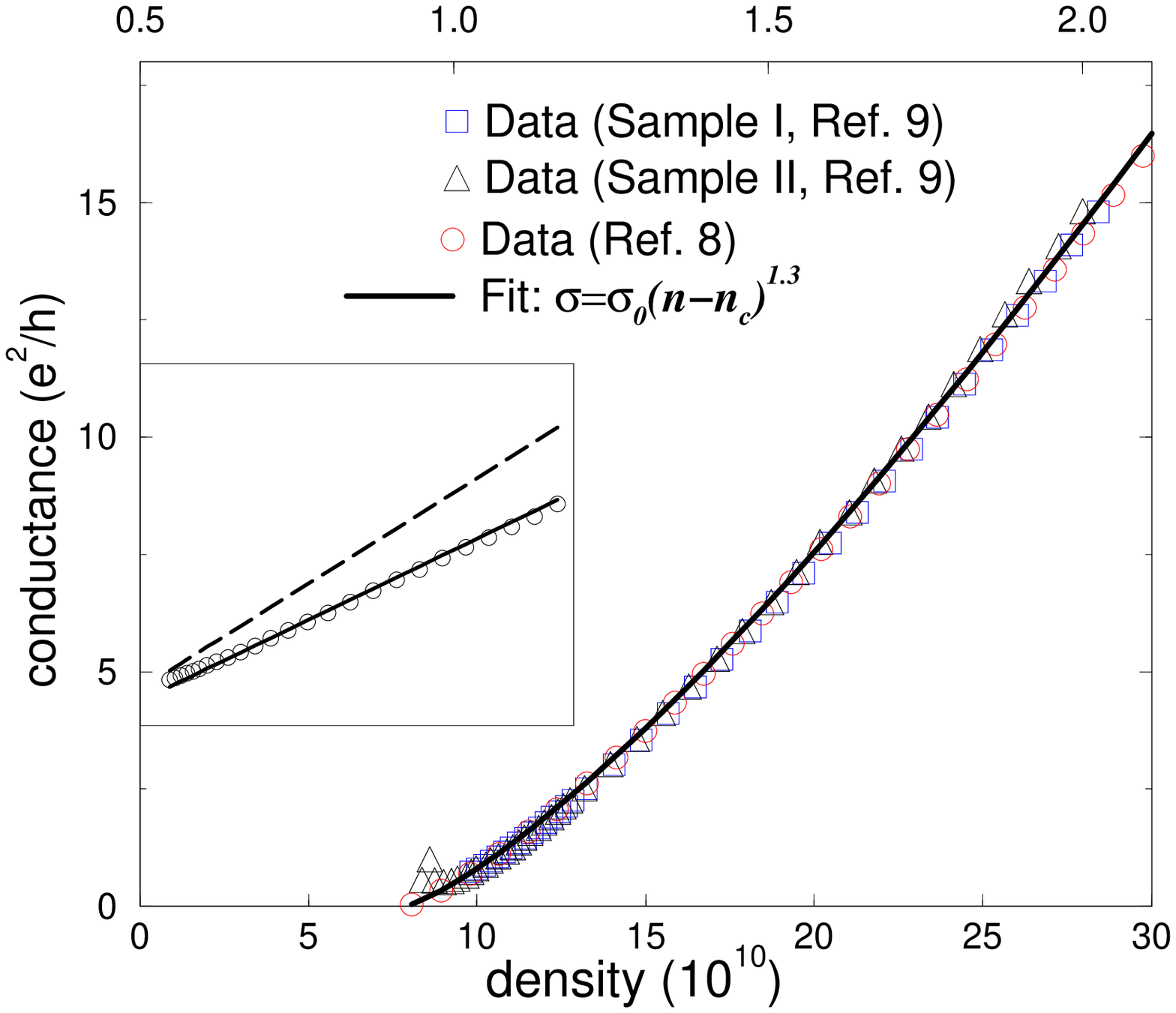}
\end{center}
\begin{small}
\vskip -0.5 truecm
Fig. 1. Comparison of the lowest temperature data of \cite{yaish}
 (two sets of data, triangles and squares,  330mK,  
 density given by the lower axis)
and of \cite{hanein} (circles,  57mK, density 
given by the upper axis)  to the prediction of 
 percolation theory (solid line). Inset: Logarithmic derivative of the data
\cite{yaish} which gives a line whose slope is the critical exponent.
The percolation prediction $(t=1.3)$ is given by the 
solid line. For comparison a $t=1$ slope is also shown (broken line).  
% Note that while the density
%scale is very different between the two experiments, the conductance scale
%is identical. 
\end{small}
\vskip 0.5 truecm

As temperature increases,  the Fermi-Dirac distribution is broadened. 
Consequently the
conductance of the insulating QPCs ($\e_c>\mu$)
increases exponentially,  while
that of the transparent ones ($\e_c<\mu$) decreases  exponentially.
Thus we expect to
see rather dramatic effects as a function of temperature. This is indeed depicted
in Fig.~2. As temperature is lowered systems with slightly different 
resistance at high temperatures will diverge exponentially with decreasing
temperatures. The resistance of systems on the metallic side ($n>n_c$) will
saturate at zero  temperature, while that of insulating   samples will 
diverge. Note that there is an upward turn even on the metallic side of the
transition. We will come back to this point below. The high-temperature 
resistance of the critical density network is naturally around  
$h/e^2$,  the only resistance scale
in this model.
% (around $2h/e^2$ in this specific geometry).
\vskip  -0.5 truecm
\begin{center}
\leavevmode \epsfxsize=3.5in
%\epsfbox{dft_fig.ps}
\epsfbox{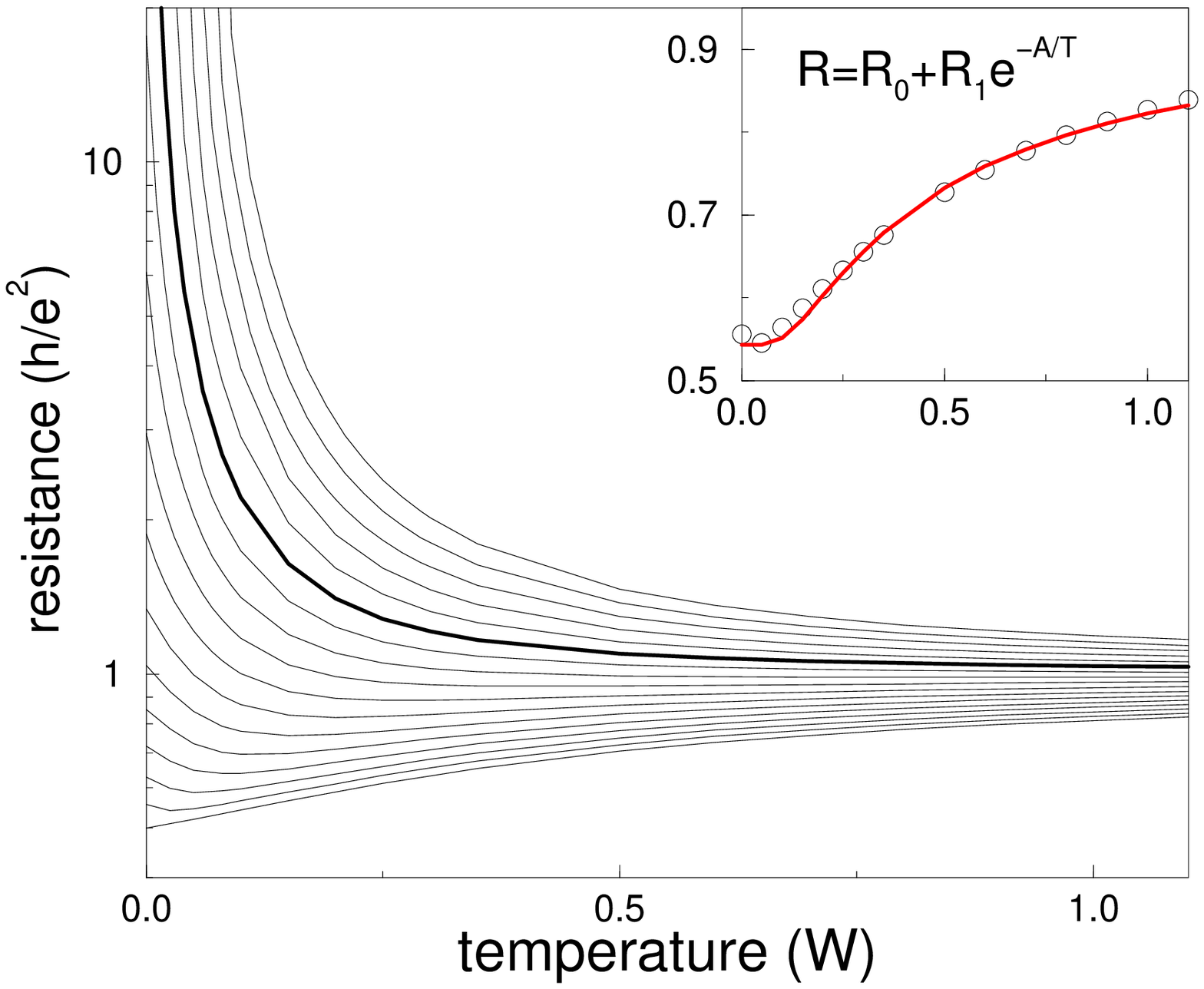}
\vskip -0.5 truecm
\end{center}
\begin{small}
Fig. 2. Temperature dependence of the resistance for systems of different
densities. Below the critical line (bold curve) all curves saturate at
zero temperature,  while above it the resistance diverges. The resistance
of the more metallic samples decreases exponentially (inset).
\end{small}
\vskip 0.5 truecm

For systems of exponentially distributed resistors
the resistance of the whole circuit is  determined by the critical
 resistor - the worse resistor in the minimal percolating network 
  \cite{ambegaokar}. Then
 the resistance of the network would be equal to the inverse of the Fermi-Dirac
 function,   $\{1+\exp[(\e_c-\mu)/kT]\} h/e^2 $,  where $\e_c$ is its
 threshold energy. Then clearly the
 overall resistance will be of the form observed experimentally,  
 $R = R_0 + R_1 \exp(-A/T)$, with $A$ varying linearly with the density and
 vanishing at the transition. In our case, the resistors
 on the metallic side have a bound distribution,  and accordingly there will
 be other resistors,  in parallel and in series,  that will contribute to
 the overall resistance of the circuit. This will not change the
 functional form,  but  will renormalize the parameters
 $R_0$, $R_1$ and $A$. Such a functional dependence on the metallic
 side is indeed found numerically and displayed in the inset. 
 In fact,  close to the transition,  on
 the metallic side,  as temperature increases, some QPCs that before
 had zero conductance,  start to conduct and add to the overall conductance.
  Since the critical percolation cluster is very ramified (in
 fact of fractal dimension),  there will be many such resistors in parallel to
 the main conducting network,  and the effect of improving these resistors will
 overcome the fact that resistors on the conducting network itself become worse. 
 This leads to a downward turn of the resistance with increasing temperature even
 on the metallic side,  the details of which may depend sensitively on the
 geometry.
  Only deeper into the metallic regime,  as seen in Fig.~2, 
  the overall  resistance increases with increasing temperature. This also
  suggests that the density at which the resistance is approximately temperature
 independent is not the true critical point,  but rather deeper on the metallic
 side. This is clearly seen in Fig.~3,  where one can see a  point  where all
 the low-temperature curves nearly cross,  well inside the metallic regime.
 The above discussion suggests that one should be cautious
 in associating the critical point with the ``temperature-independent'' point, 
 as done routinely in the experiment interpretations.
\vskip -0.5 truecm
\begin{center}
\leavevmode \epsfxsize=3.5in
%\epsfbox{dft_fig.ps}
\epsfbox{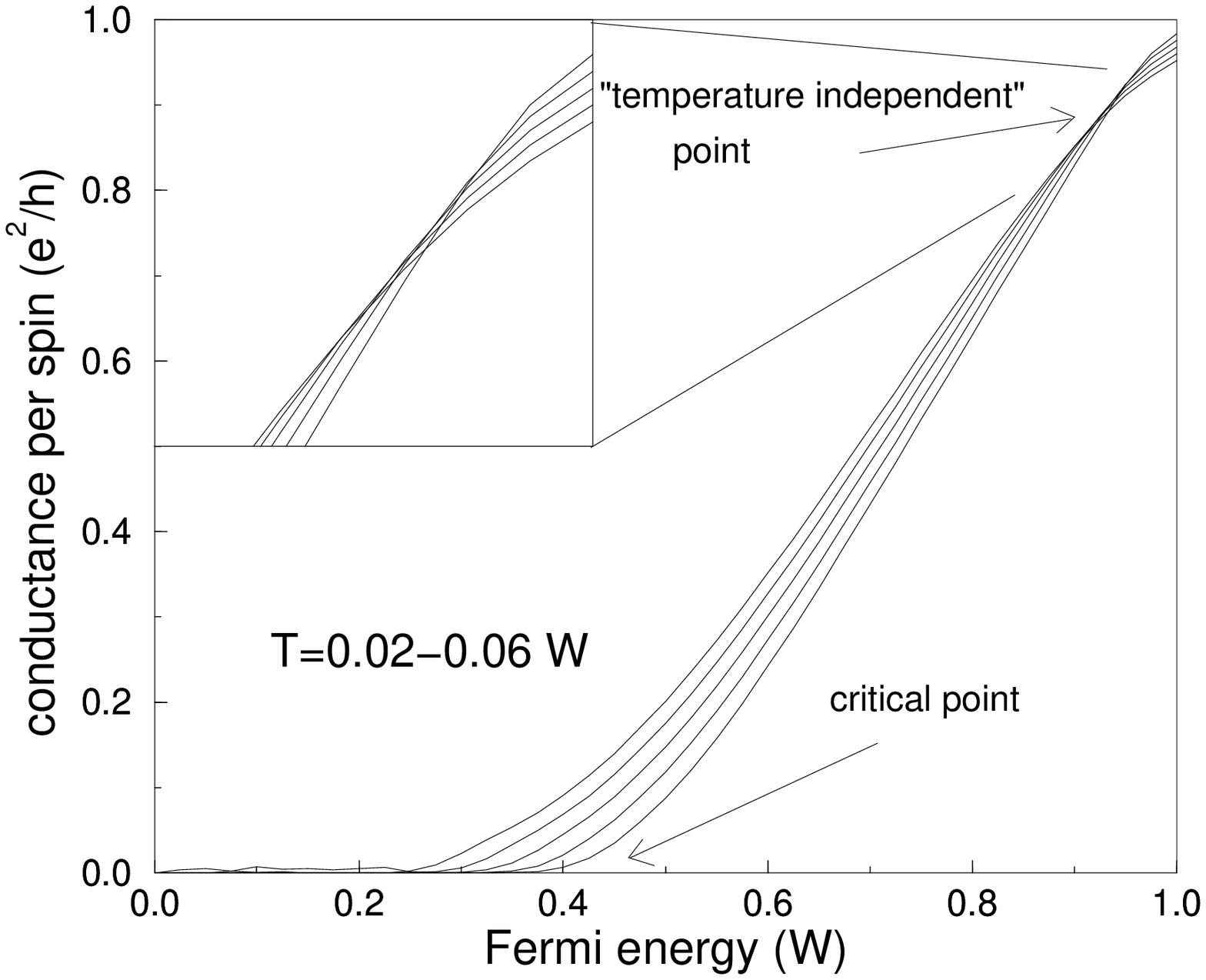}
\end{center}
\vskip -0.5 truecm
\begin{small}
Fig. 3. Conductance vs. density (Fermi energy) for several temperatures. 
There is a density, well above the true critical point,
 where the curves seem to cross each other. 
\end{small}
\vskip 0.5 truecm
We turn next to the effects of a magnetic field. The effect of a
 parallel  field is straightforward to understand, as there have
 been several studies of transport through a QPC in parallel fields
 \cite{qpc_parallel}. These experimental and theoretical
  studies demonstrated that the 
 threshold density  where the QPC opens up {\sl
 increases} parabolically with the in-plane magnetic field. This effect was
 attributed to  coupling of
 the in-plane motion to the strong confinement in the vertical
 direction,  leading to an increase in the energy levels. Since this increase
 occurs for all QPCs in the sample,  such
 a field in our case will strongly inhibit the metallic behavior:
 QPCs which
 were conducting at zero field,  will have exponentially small conductance
 with increasing field. Once the density of conducting QPCs falls
 below the critical density,  the system becomes an insulator,  in agreement
 with experimental observations.
 
The situation in perpendicular magnetic fields is more interesting,  as
QH states are formed.
Transport  through a single QPC in perpendicular field and
the crossover between the zero field limit and the QH limit have
been studied in detail \cite{vanwees}. As expected, one finds that the critical
energy oscillates with magnetic field due to the depopulation of Landau
levels. In our case,  we expect the oscillations to be smoothed out by the
disorder and by the averaging over many QPCs. Thus only the strongest
oscillation,  near  $\nu=1$,  may survive,  leading to a single dip in the
critical density vs. magnetic field plot,  as was observed experimentally. 
This is indeed in agreement with our numerical calculation. We studied the energy
levels of one puddle of electrons,  which we modeled by a circular disk,  in the
presence of disorder \cite{disorder}. In Fig.~4 we plot the ``critical density''
 -- the number of electrons
that need to occupy the puddle,  so that the energy of the highest-energy 
electron will be enough to transverse the QPC \cite{fertig}, 
equivalent in the bulk system to the critical density -- as a function of 
magnetic field. Indeed we see a dip near $\nu=1$ 
%(the dip actually shifts away
%from that value of magnetic field with increasing disorder),  
with all other
oscillations smoothed out by the disorder. This curve has a strong resemblance
to the experimental data \cite{magfield} (inset). 
In addition,  we expect that as the
magnetic field is lowered below the $\nu=1$ minimum, more than one channel
will transverse  some  QPCs,  leading to an increase in the
critical conductance,  as indeed reported experimentally. 
\vskip -0.5 truecm
\begin{center}
\leavevmode \epsfxsize=3.5in
%\epsfbox{dft_fig.ps}
\epsfbox{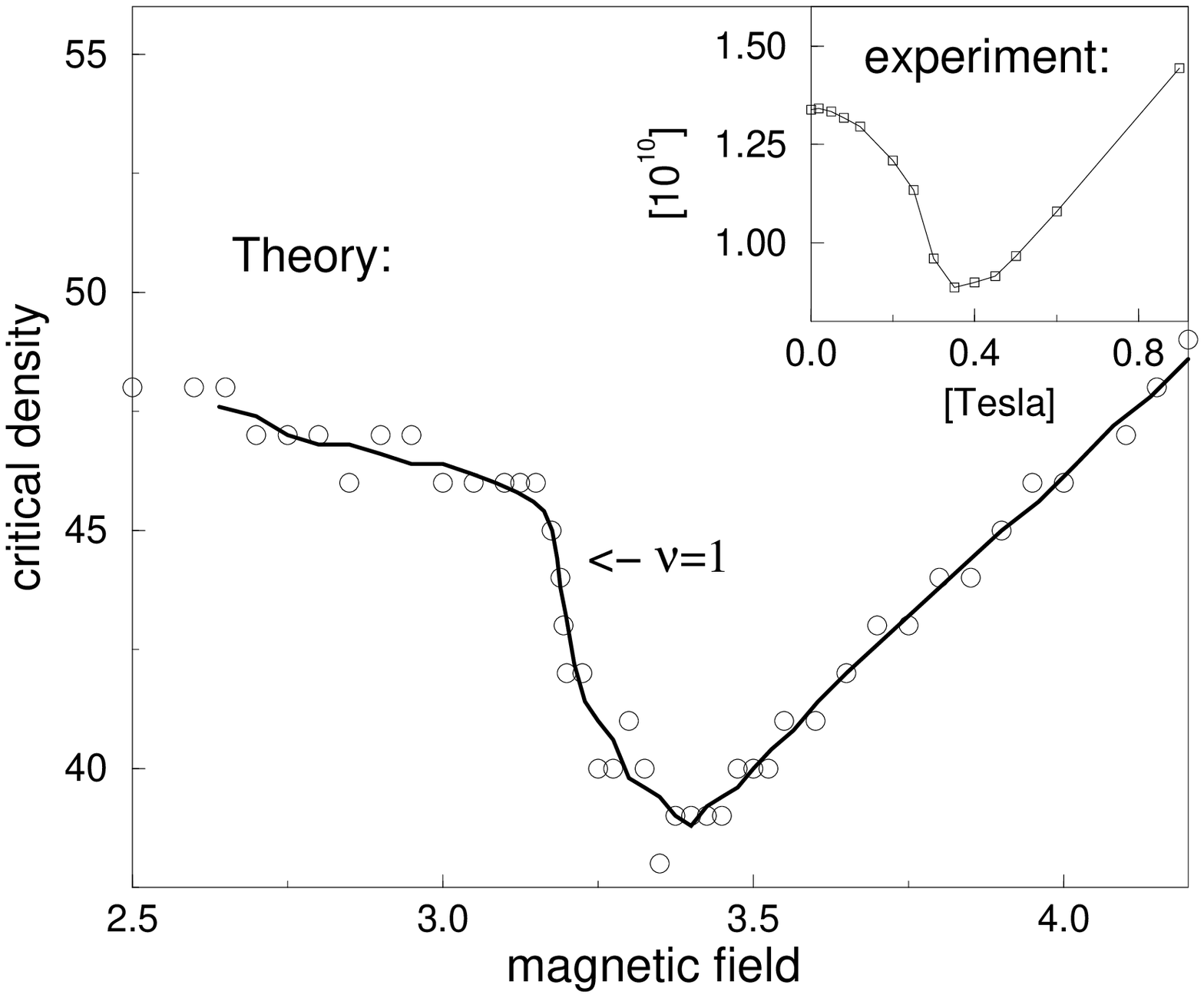}
\end{center}
\vskip -0.5 truecm
\begin{small}
Fig. 4. The critical density - the number of electrons in the puddle,  so that
the topmost energy will allow transport through the point contact - as a
 function of magnetic field,  in the presence of a finite disorder. The 
continuous curve is an averaged fit through the (necessarily integer) data
points. Inset: the corresponding experimental data \cite{magfield}.
\end{small}
\vskip 0.5 truecm

All the above results and discussion demonstrated that many of the experimental
observations can be explained in the context of the simple model introduced
here. Nevertheless there are clearly other physical effects that need to be
included in order to have a full picture of the experiments. In particular, 
electron-electron interactions are expected to play an important role in
these low densities. As we can regard the metallic puddles described above as
quantum dots,  one can use the abundant information about the role of
interactions in such structures \cite{qdots}, to gain additional understanding
of the characteristics of the puddles and the phase separation.
Other effects, 
including the energy dependence of the transmission
coefficient and the possibility of more than one channel through the QPCs, 
the temperature dependence of the dephasing length,  
the role of interband-scattering \cite{yaish} and temperature-dependent 
impurities
\cite{altshuler} may also be important to understand quantitative aspects of
 the data. Nevertheless,  the fact that several important aspects of the
 experimental data can be explained in the context of a simple model is quite
  encouraging. We expect that the model presented here apply mostly 
 to the  GaAs samples.
  
The picture described above can be checked experimentally. The experiments
verifying the percolative structure in the QH regime \cite{qh} can be 
extended to the zero-field systems. (An experimental evidence for phase
separation was observed at zero field in \cite{israel}.) An even more direct 
evidence
of the percolative nature of the system will be local probes \cite{ashoori}.
In fact, an enhancement in the fluctuations of the local chemical potential
has already been observed \cite{amir} as the system enters the ``insulating'' 
phase,  in a similar fashion to the enhancement of chemical potential 
fluctuations with closing of the barriers forming a quantum dot \cite{udi}.
In fact,  a ``smoking gun'' verification of the picture presented here,  will
be periodic oscillations of the local chemical potential on the insulating side, 
due to depopulation of the Landau levels,  as was observed in quantum dots
 \cite{paul}.

 I thank many of my colleagues for fruitful discussions:
 A.~Auerbach, Y.~Gefen, Y.~Hanein, D.~Shahar, E.~Shimshoni,  U.~Sivan, 
 A.~Stern,   A.~Yacoby and  Y. Yaish. In particular, I would like to 
thank Y. Hanein \& D. Shahar and U. Sivan \& Y. Yaish for making their
data available to me. This work
  was supported by THE ISRAEL SCIENCE FOUNDATION  - Centers of Excellence
 Program,  and by the German Ministry of Science.

\end{multicols}

\begin{references}
\bibitem{disorder}
To simlify the numerics,  we chose random potentials with azimuthal symmetry.
\end{references}
\end{document}